\documentclass[aps, prd, amssymb, showpacs, floatfix, twocolumn, superscriptaddress, nofootinbib]{revtex4-2}
\usepackage{amsmath,amsfonts,graphics,epsfig}
\usepackage[cp1251]{inputenc}
\begin{document}


\title{Relativistic composite-particle theory of the gravitational form factors of pion:
quantitative results}

\author{A.F.~Krutov}
\email{a\underline{$\hphantom{z}$}krutov@rambler.ru}, \affiliation{Samara State Technical University, 443100 Samara,
        Russia},\affiliation{Samara Branch, P.N.~Lebedev Physical Institute of the Russian Academy of Sciences, 443011 Samara, Russia},
\author{V.E.~Troitsky} \email{troitsky@theory.sinp.msu.ru}
\affiliation{D.V.~Skobeltsyn
        Institute of Nuclear Physics,\\
        M.V.~Lomonosov Moscow State University, Moscow 119991, Russia}
\date{\today}
\begin{abstract}
We use a version   of the instant-form  relativistic quantum mechanics
of composite systems to obtain the gravitational form factors of the pion
in a common approach to its electroweak and gravitational properties. In
the preceding work [Phys.Rev. D \textbf{103},  014029 (2021)] we
formulated the mathematical background, presented the principal
scheme of calculation and testified the obtained qualitative results  to
satisfy the general constraints given by the principles of the theory of
hadron structure. In the present work we give the detailed calculation
of the gravitational form factors in large range of momentum transfer,
their static limits and the slopes at zero value, the mean-square
mass and mechanical radii of the pion. Now we take into account the qravitational
structure of the
constituent quarks. We  show that the results are almost insensitive to
the type of the model two-quark wave function in a close analogy to the
case of the pion electromagnetic form factor. We present a correct
calculation of the form factor $D$ and corresponding
matrix element of the energy-momentum tensor, going beyond the scope of
the modified impulse approximation. Most of the parameters that we use for
the calculation had been fixed even earlier in our works on the pion
electromagnetic form factors. The only free parameter is the $D$-term of
the constituent quark, which we fix  by fitting the result for the slope
at zero of the normalized to pion $D$-term  form factor $D$ of pion, to a
choosen experimental value.
\end{abstract}

\maketitle

 \section{Introduction}

The understanding of the gravitational structure of hadrons
is a fundamental problem of particle  physics. To consider this problem
one needs to involve specific mathematical objects: the gravitational form
factors (GFFs) of hadron (and in general the energy-momentum tensor
(EMT)). These functions are in the focus of investigation in numerous
recent works (see, e.g., the reviews \cite{LeL14, Ter16, PoS18, BuE18,
CoL20}). The mathematical background of these investigations  was laid in
the sixties of the last century  \cite{KoO63, ChS63, Pag66} but becomes
of large use only during the last decade.

It is the formalism of Ref.\cite{ChS63} that we use in our investigations.
In the preceding work \cite{KrT21} we formulated
the mathematical statements and described
the principal scheme of the method; we testified that the derived
gravitational characteristics of the pion  satisfy the constraints
given by the general principles of the theory of hadron
structure. In the present work we refine the approach   as to give a
detailed technique of calculation, thus transforming it into a
quantitative method. The refinement includes three main points. First, we
take into account the gravitational structure of the constituent quarks.
Second, we analyze different types of the two-quark wave functions in the
pion. Third, we formulate a minimal way to overstep our
modified impulse approximation (MIA) for correct description of the pion
form factor $D$.
Note that the pion EMT contains two
form factors, and the second form factor $A$ is described well in the
frame of MIA.

The majority of papers concerning closely the problem under
consideration were reviewed  in \cite{KrT21} and we do not
repeat the review in the present paper. Here we discuss only some recent
results. Although the papers published during last year are dealing
mainly with the proton, they show the modern trends and perspectives
in general.
At present, there is no possibility to obtain
directly the data on gravitational characteristics, including the
$D$-term. The information about the GFFs is  usually extracted from the
hard-exclusive processes described in terms of unpolarized generalized
parton distribution (GPD) \cite{Pol03, PoS19}. The GPD gives the
information on the space distribution of strong forces that act on quarks
and gluons inside hadrons. In this connection, in works \cite{DuL21,
GuJ21, BuE21, ChN21} the processes of deeply virtual Compton scattering
(DVCS) on nucleons were investigated. In the paper \cite{DuL21} new
technique of artificial neural networks was used for the reduction of
model dependence of the data handling. The distribution of shear forces
inside the proton was obtained for the first time in \cite{BuE21} and also
the proton form factor $D$ was constructed by fitting of three
parameters in the multipole-type decomposition. The kinematic
corrections to the cross-section appearing as a consequence of
non-uniqueness of description of the photon in final state
on the light cone, is discussed in the work \cite{GuJ21}. It is shown that
these corrections can be significant in the kinematic regions of future
experiments. The experimental study of DVCS is realized in time-like
region for the first time in \cite{ChN21}.

In another set of works (Refs. \cite{WaK21, KoW21, WaK21PRD, Kha21,
MaZ21}), the gravitational characteristics of hadrons are extracted from
the data on near-threshold vector mesons photoproduction. In the study
\cite{WaK21}, the mass radius is calculated not only for the proton, but
for the deuteron, too.
The analyzis of the possible sources
of, systematically obtained in numerous recent calculations, difference
between charge and mass radii of the  hadrons is given in \cite{Kha21}.

The low-energy chiral effective field theory(chiral EFT), was used in
\cite{EpG21} to calculate the $\rho$-meson GFFs and in \cite{GeP21} to
describe the mechanical stresses inside the nucleon.

Constituent quark models are exploited in the papers
\cite{TaL21, ChM21, LoS22}. In particular, the spin-orbital correlations
in pion in a quark model on the light cone are studied in \cite{TaL21} in
terms of GPD and of generalized transverse momentum distributions. In
\cite{ChM21}, the proton GFFs are derived in a light-front quark-diquark
model. The authors of \cite{LoS22} study general problems of the
interpretation of the experimental data and theoretical methods of
calculation for nucleon GGFs. The investigation is focused on space
distributions of the energy, shear stresses, momentums, angular momentums
in different frame systems. The authors use the quark bag  model in
large-$N_c$ limit in various cases - from non-relativistic to
ultrarelativistic.

As to non-hadronic systems, the form factor $D$, the distribution of
matter, and internal stresses in the electron are obtained in \cite{MeP21}
in the one-loop approximation of QED.

It seems obvious that  because of extreme weakness of the
gravitational interaction at hadron scale, the information on  GFFs will
be extracted from electroweak processes as yet. So, a theory which
describes electroweak and gravitational properties simultaneously,
based on the unique foundations, and uses common model parameters, is
welcome. The approach that we present here possesses these features.

We use a particular variant of the instant-form Dirac
\cite{Dir49} relativistic quantum mechanics (RQM) (see, e.g.,
\cite{LeS78, KeP91, Coe92, KrT09}) extended for composite systems. The
approach was successfully used to describe the pion electromagnetic  form
factor (see., e.g., \cite{KrT01, KrT02, KrT03}). We have shown
\cite{KrT21} that the pion GFFs can be derived in the same formalism using
the same approximations and the same model parameters, adding only one new
parameter fixed by fitting  the slope at zero of the normalized to pion
$D$-term  form factor $D$ of pion.

In the  present paper, we take into account the gravitational structure of
the constituent quarks and different forms of the quark-antiquark wave
function. It is worth noting that the including of the gravitational
structure of costituents in our relativistic model means an implicit
accounting for gluons. Their degrees of freedom are incorporated in the
parameters describing constituent quarks.

Let us emphasize that it was just the account of the structure
of the constituents \cite{KrT01} that made it possible to predict the
behavior of the pion charge form factor at intermediate and high momentum
transfers. We have shown that, when two model parameters were fixed by the
charge mean square radius and the lepton decay constant, then the form
factors only weakly depended on the choice of model interaction of
quarks in the pion. The curves corresponding to different interactions but
one and the same value of quark mass were agglomerated into narrow groups,
or bunches. The  chosen group of curves with the constituent mass $M=$0.22
GeV had predicted, with surprising accuracy, the values of the pion charge
form factor which were measured a decade later in JLab experiments. By the
way, this value of the constituent-quark mass has been admitted and
confirmed by other authors (from the well known work \cite{GoI85} to
recent result \cite{HaA21}).

Let us list some other advantages of our model with fixed parameters.
The obtained pion charge form factor at high momentum transfer coincides
with the QCD predictions in the
ultraviolet limit, reproducing correctly not only the
functional form of the QCD asymptotics, but also the numerical
coefficient~\cite{KrT98, TrT13, KrT17} (analogous results
were obtained for the kaon \cite{TrT21}). The method allows for an analytic
continuation of the pion electromagnetic form factor from the space-like
region to the complex plane of momentum transfer squares and gives an
adequate description of the pion form factor in the time-like region
\cite{KrN13}. Interesting results of the approach in the case of its
generalization to vector mesons, which required to add the anomalous
magnetic moment of quark, were obtained for the $\rho$ meson \cite{KrP16,
KrP18}.

As was mentioned above, we need to
overstep the frame of MIA for calculating correctly the
pion form factor $D$. Here we present a possible minimal extension based
on the non-relativistic limit of MIA. In the relativistic series expansion
of the ill-defined terms of the form factor we preserve the main
contributions only.

In the present paper we extend our model with non-point-like quarks to
obtain the pion GFFs while making an essential assumption, which looks
natural. We suppose that the functional forms of electroweak
and gravitational form factors of constituent quark are the same.
Also we suppose that charge and mass radii of quark, and the slope at zero
of the normalized form factor $D$, - all these three values are equal.
From one hand, this gives an opportunity to add to the model
only one new parameter. From the other hand, the results of calculation
are rather reasonable and plausible.

We fix the new free parameter,
the quark $D$-term (instead of quark anomalous magnetic moment
in vector meson case of electroweak processes) by
fitting  the slope at zero of the normalized to pion $D$-term  form factor
$D$ of pion
to the value obtained
in Ref. \cite{KuS18}. The scheme of narrow bunches of curves,
corresponding to different wave functions, works in the case of pion GFFs,
too. As we show below it is possible to choose the quark $D$-term as the
main characteristic of the bunch and so derive the value of the free
parameter.

Other model parameters, which describe the mass and
the gravitational structure of the constituent quarks and the quark
interaction in the pion, are directly transported from our electroweak
model. It is important to emphasize that the interval of the values of the
quark $D$-term obtained by this fixation gives the value of
the slope at zero of the normalized form factor $D$ of pion
\cite{KuS18} for any quark interaction (for each of three two-quark wave
functions) without additional variation of the previous values of the
parameters.

This set of parameters -- those inherited from the electroweak calculation
plus the quark $D$-term --  allow  us to describe all other gravitational
characteristics of the pion: the GFFs in a large range of momentum
transfers, their static limits, including the pion $D$-term and the mean
square mass radius. So, the results of calculations of the
present paper can be considered as firm predictions in the following
sense. If a value of, for example,
the slope at zero of the normalized to $D$-term form factor $D$ of pion
extracted
from a  future experiment, is admitted by the community of specialists,
then other gravitational characteristics can be predicted following our
prescription. The success demonstrated previously by our method inspires
hope on its validity in describing the pion GFFs  that can be obtained in
future experiments.

Let us remind that the first estimations of the pion GFFs were
extracted from the data of Belle collaboration program KEKB
\cite{KuS18, Mas16}. Today, the real perspective of GPD estimation for
light mesons, including the pion, is connected with the future experiments
on the SuperKEKB collider.
In general, we look forward for the
experiments at the electron-ion collider (EIC) \cite{Acc, Abd}, Chinese
electron-ion collider (EIcC) \cite{Che18, And21}, and Large hadron-electron
collider (LHeC) \cite{Abe12}.

Moreover, it is possible that our unified approach  can construct another
bridge between electroweak and gravitational properties of composite
systems, complementary to GPD.

The rest of the paper is organized as follows.
In Sect.\ref{sec: Sec 2} we present the main basic points of IF RQM  and
the equations for the pion GFFs. Then in Sect.
\ref{sec: Sec 3} we describe the gravitational structure of the
constituent quarks and present the details of calculation.  In
Sect.\ref{sec: Sec 4} we give the results of the calculation of the pion
GFFs up to 10\,GeV$^2$, their static limits and the mean square mass and
mechanical radii. We briefly discuss the results and conclude in
Sect.\ref{sec: Sec 5}.

\section{Main basic points of IF RQM and the calculation of the pion GFFs}
\label{sec: Sec 2}

Let us recall briefly some important features of the IF RQM algebraic
structure. The basic point is the direct realization
of the Poincar\'e algebra on the set of dynamical
observables of  a composite system (see the
reviews \cite{LeS78, KeP91, Coe92, KrT09}).
In this context, the fundamental property of  the
Poincare algebra, as compared to the algebra of the Galilean group,
is the following. The adding of the operator of constituent interaction
to the total energy operator (that is to zero component of total
momentum) requires the including of the interaction also in the
operators of other observables to preserve the algebraic structure.
Different forms of relativistic Dirac dynamics correspond to different
ways of realizing the interaction including, which are characterized by
different kinematic subgroups, namely subgroups of interaction-independent
observables. The kinematic subgroup in the case of IF RQM contains
rotations and translations of three-dimensional space.
From the point of view of the principles underlying the RQM theory, it
occupies an intermediate position between local quantum field theory
and non-relativistic quantum mechanical models. In particular,
constituents of composite system are assumed to lie on the mass
shell, and corresponding wave function is defined as eigenfunction of the
complete set of commuting operators.  In the case of IF RQM this set is:
\begin{equation}
	{\hat M}_I^2\;(\hbox{or}\;\hat M_I)\;,\quad
	{\hat J}^2\;,\quad \hat J_3\;,\quad \hat {\vec P}\;,
	\label{complete}
\end{equation}
here $\hat M_I$ the mass operator for the
system with interaction, ${\hat J}^2$ is the operator of the square of the total angular
moment, $\hat J_3$ is the operator of the projection of the total
angular moment on the $z$ axis and $\hat {\vec P}$ is the operator of the
total momentum.

In the IF RQM the operators ${\hat J}^2\;,\; \hat J_3\;,\; \hat {\vec P}$
coincide with corresponding operators for the composite system without
interaction and only the term $\hat M_I^2\;(\hat M_I) $
is interaction depending.

To solve the eigenfunction problem for the set (\ref{complete}) it is
necessary to choose an appropriate basis in the Hilbert space of the state
of the composite system. In the case of system of two
constituent quarks one can use, first, the basis of individual spins and
momenta (see \cite{KrT21} for details):
\begin{equation}
	|\,\vec p_1\,,m_1;\,\vec p_2\,,m_2\,\!\rangle  =
	|\,\vec p_1\,m_1\,\!\rangle \otimes
	|\, \vec p_2\,m_2\,\rangle\;,
	\label{p1p2}
\end{equation}
where $\vec p_1,\,\vec p_2$ are the 3-momenta of particles,
$m_1,\,m_2$ are the projections of spins to the $z$ axis.

Second, it is possible to use the basis in which the
motion of the center  of mass of two particles is separated:
\begin{equation}
	|\,\vec P,\;\sqrt {s},\;J,\;l,\;S,\;m_J\,\rangle\;,
	\label{Pk}
\end{equation}
where $P_\mu = (p_1 +p_2)_\mu$, $P^2_\mu = s$, $\sqrt {s}$
is the invariant mass of the  system of two
particles, $l$ is the orbital momentum in the center-of-mass system
(c.m.s.), $\vec S\,^2=(\vec S_1 + \vec S_2)^2 = S(S+1)\;,\;S$ is the
total spin in c.m.s., $J$
is the total angular momentum, $m_J$ is the projection of the total
angular momentum.

The bases (\ref{p1p2}) and (\ref{Pk}) are linked
by the Clebsch-Gordan decomposition of a direct
product (\ref{p1p2})  of two irreducible representations of the Poincar\'e
group into irreducible representations (\ref{Pk}) \cite{KrT09}.

In the basis (\ref{Pk})  three out of four operators in the complete set
(\ref{complete}) (except $\hat M_I$) are diagonal. So, the two-quark
wave function for the pion in the basis (\ref{Pk}) has the following form:
\begin{equation}
\langle\vec P\,,\,\sqrt {s}|\vec p_\pi\rangle  =
N_C\,\delta (\vec P\, - \vec p_\pi)\,\varphi(s)\;,
	\label{wfI}
\end{equation}
where $\vec p_\pi$ is the pion 3-mometum. Here we do not exploit the
implicit form of the normalization constant $N_C$ that can be found in
\cite{KrT21}. The zero values of pion quantum numbers are omitted in the
notation of basis vectors (\ref{Pk}).

The wave function of intrinsic motion is the eigenfunction of the operator
$\hat M_I^2\;(\hat M_I)$ and in the case of two particles of equal masses
is (see, e.g., \cite{KrT02})
$$
	\varphi(s(k)) = \sqrt[4]{s}\,k\,u(k)\;,\quad s = 4(k^2 + M^2)\;,
$$
\begin{equation}
	\int\,u^2(k)\,k^2\,dk = 1\;,
	\label{phi}
\end{equation}
where $u(k)$ is a model quark-antiquark wave function of the pion and
$M$ is the mass of the constituent quarks.

Let us construct now the pion EMT in IF RQM. Using the general
method of the relativistic invariant parametrization of the matrix
elements of the local operators we have obtained in \cite{KrT21} the
followig form:
$$
\langle \vec p_\pi\left|T^{(\pi)}_{\mu\nu}(0)\right|\vec p_\pi\,'\rangle = \frac{1}{2}G^{(\pi)}_{10}(t)K'_\mu K'_\nu +
$$
\begin{equation}
	+ G^{(\pi)}_{60}(t)\left[tg_{\mu\nu} - K_\mu K_\nu\right]\;,
	\label{Tpi}
\end{equation}
where $G^{(\pi)}_{10}, G^{(\pi)}_{60}$ are gravitational form factors,
$g_{\mu\nu}$ is the metric pseudotensor and
$$
K_\mu = (p_\pi - p_\pi')_\mu\,,\quad K'_\mu = (p_\pi + p_\pi')_\mu \,.
$$

We present the decomposition of the l.h.s. of (\ref{Tpi}) in terms of the
basis (\ref{Pk}) as a superposition of the same tensors as in the r.h.s.
of (\ref{Tpi}), and so obtain (see \cite{KrT21} for details) the pion
gravitational form factors in the following form of the functionals,
given on two-quark wave functions (\ref{wfI}), (\ref{phi}):
$$
G^{(\pi)}_{i0}(t) =
\int\,d\sqrt{s}\,d\sqrt{s'}\,
\varphi(s)\tilde G_{i0}(s,t,s')\varphi(s')\;,
$$
\begin{equation}
	i=1,6\;.
	\label{int ds=Gpi}
\end{equation}
here $\tilde G_{i0}(s,t,s'),\,i=1,6$ are the
Lorentz-invariant regular distributions.

To calculate the invariant distributions in r.h.s. of (\ref{int ds=Gpi})
one can use MIA. Let us discuss this
problem in more detail. Consider the commonly used standard impulse
approximation (IA). In general, the EMT of a composite system has the
following form \cite{KrT21}:
\begin{equation}
	T = \sum_{k}\,T^{(k)} +
	\sum_{k<m}\,T^{(km)}+\ldots\;,
	\label{T=Tk+Tkm}
\end{equation}
where the first term presents the sum of one-particle EMTs, the second
term presents the sum of two-particle EMT, and so on. The first sum
describes the scattering of a projectile by each independent
constituent, the second sum describes the scattering by two constituents
simultaneously and so on. The standard IA leaves in (\ref{T=Tk+Tkm})  only
the first term:
\begin{equation}
	T \approx \sum_{k}\,T^{(k)}\;.
	\label{IA}
\end{equation}
To construct the pion GFFs we use
a modified impulse approximation that we first formulated earlier (see,
e.g., Refs. \cite{KrT02,KrT03} and the review \cite{KrT09}) In contrast to
the baseline impulse approximation, MIA is formulated in terms of the
reduced matrix elements, that is form factors, and not in terms of the
operators itself. So, in MIA there appears important objects -- the
free gravitational form factors presenting the gravitational
characteristics of systems without interaction.

Consider the system of two free constituent quarks \cite{KrT21}.
Note, that in the work \cite{HuS18} it was shown that the form factor $D$
is zero in the case of point-like free fermions. In contrast, our
constituent quarks have all properties of realistic particles with
internal structure that is described by a set of form factors including
form factor $D$.

The corresponding
relation based on  the Clebsch-Gordan decomposition of the EMT of a system
of non-interacting fermions with total quantum numbers of pion
$J=l=S=0$ has the following form  \cite{KrT21}:
$$
\langle P,\sqrt{s}\left|T^{(0)}_{\mu\nu}(0)\right|P',\sqrt{s'}\rangle =
$$
$$
= \sum\int\frac{d\vec p_1}{2p_{10}}\frac{d\vec p_2}{2p_{20}}
\frac{d\vec p\,'_1}{2p'_{10}}\frac{d\vec p\,'_2}{2p'_{20}}
\langle P,\sqrt{s}\left|\right.\vec p_1,m_1;\vec p_2,m_2\rangle\times
$$
$$
\left[\langle \vec p_1,m_1\left|\right.\vec p\,'_1,m'_1\rangle
\langle p_2,m_2\left|T^{(u)}_{\mu\nu}(0)\right|p'_2,m'_2\rangle\right. +
$$
$$
+ \left.\langle \vec p_2,m_2\left|\right.\vec p\,'_2,m'_2\rangle
\langle p_1,m_1\left|T^{(\bar d)}_{\mu\nu}(0)\right|p'_1,m'_1\rangle\right]\times
$$
\begin{equation}
	\langle \vec p\,'_1,m'_1;\vec p\,'_2,m'_2\left|\right.
	P',\sqrt{s'}\rangle\;,
	\label{T0p1p2}
\end{equation}
where $\langle P,\sqrt{s}\left|\right.\vec p_1,m_1;\vec p_2,m_2\rangle$ is
the Clebsch-Gordan coefficient,  the sums are over the variables
$m_1,\,m_2,\;m'_1,\,m'_2$. In l.h.s. zero discrete quantum numbers in state
vectors are ignored.

Using the general method of parametrization of local-operators matrix
elements \cite{ChS63,KrT21} we write the matrix element in l.h.s. as:
$$
\langle P,\sqrt{s}\left|T^{(0)}_{\mu\nu}(0)\right|P',\sqrt{s'}\rangle =
$$
$$
= \frac{1}{2}G^{(0)}_{10}(s,t,s')A'_\mu A'_\nu +
$$
\begin{equation}
	+ G^{(0)}_{60}(s,t,s')\left[t\,g_{\mu\nu} - A_\mu A_\nu\right]\;,
	\label{T0}
\end{equation}
where $G^{(0)}_{i0}(s,t,s'),\,i=1,6$ are free two-particle GFFs,
$$
A_\mu = \left(P - P'\right)_\mu\;,\quad A^2 = t\;,
$$
$$
A'_\mu = \frac{1}{(-t)}\left[(s - s' - t)P_\mu + (s' - s -
t)P'_\mu\right]\;.
$$

The method gives for the one-particle matrix elements in
r.h.s. of (\ref{T0p1p2}) the form:
$$
\langle p,m\left|T^{(q)}_{\mu\nu}(0)\right|p',m'\rangle =
\sum_{m''}\langle m\left|D_w^{1/2}(p,p')\right|m''\rangle\times
$$
$$
\langle m''\left|(1/2)g^{(q)}_{10}(t)K\,'_\mu K'_\nu
+ ig^{(q)}_{40}(t)\left[K'_\mu\,R_\nu + R_\mu\,K'_\nu\right] +\right.
$$
\begin{equation}
	\left. + g^{(q)}_{60}(t)\left[tg_{\mu\nu} - K_\mu K_\nu\right]
	\right|m'\rangle \;,
	\label{Tq}
\end{equation}
$q = u,\,\bar d$, $D_w^j(p,\,p')$ is the transformation operator from the
small group, the matrix of three-dimensional rotation,
$g^{(u,\bar d)}_{i0},\,i=1,4,6$ are the constituent-quark GFFs. Their
links with conventional notations is given later in Sect.\ref{sec: Sec 3}.
$$
K_\mu = (p - p')_\mu\,,\quad K'_\mu = (p + p')_\mu \,,
$$
\begin{equation}
	R_\mu = \epsilon _{\mu \,\nu \,\lambda
		\,\rho}\, p^\nu \,p'\,^\lambda \,\Gamma^\rho (p')\;.
	\label{kk'RG}
\end{equation}
Here $\Gamma^\rho (p')$ is well known 4-vector of spin (see, e.g.,
\cite{ChS63, KrT21, KrT09, KrT03}), $\epsilon _{\mu \,\nu
	\,\lambda\,\rho}$ is the absolutely antisymmetric pseudotensor of rank 4,
$\epsilon_{0\,1\,2\,3}= -1$.

Substituting (\ref{T0}), (\ref{Tq}) into (\ref{T0p1p2}), multiplying both
sides by scalars, first, by $A'_\mu A'_\nu$, and second by
$g_{\mu\nu}$, performing  the integrations  and summations, we obtain the
system of equations for the free two-particle form factors  which enter
(\ref{T0}). MIA means replacing the invariant distributions in
r.h.s. of (\ref{int ds=Gpi}) with free two-particle form factors
from the equations (\ref{T0p1p2}), (\ref{T0}). The physical meaning of MIA
is equivalent to that of the universally accepted IA
(\ref{IA}) because the free two-particle form factors are given in terms of
one-particle currents (\ref{T0p1p2}), (\ref{Tq}). So, we obtain  the
expressions for the pion gravitational form factors in MIA.  For
convenience, we present them in a slightly different form than in
\cite{KrT21}:
$$
G^{(\pi)}_{10}(t) =
\frac{1}{2}\left[g^{(u)}_{10}(t)+g^{(\bar d)}_{10}(t)\right]\,G^{(\pi)}_{110}(t) +
$$
\begin{equation}
	+ \left[g^{(u)}_{40}(t)+g^{(\bar d)}_{40}(t)\right]\,G^{(\pi)}_{140}(t)\;,
	\label{Gpi1}
\end{equation}
$$
G^{(\pi)}_{60}(t) =
\frac{1}{2}\left[g^{(u)}_{10}(t)+g^{(\bar d)}_{10}(t)\right]\,G^{(\pi)}_{610}(t) +
$$
$$
+ \left[g^{(u)}_{40}(t)+g^{(\bar d)}_{40}(t)\right]\,G^{(\pi)}_{640}(t)  +
$$
\begin{equation}
	+ \left[g^{(u)}_{60}(t)+g^{(\bar d)}_{60}(t)\right]\,G^{(\pi)}_{660}(t)\;.
	\label{Gpi6}
\end{equation}
Here $g^{(q)}_{i0}(t)\,,\,q=u,\bar d\,,\,i=1,4,6$  are the GFFs of the
constituent quarks, also introduced previously in (\ref{Tq}).

To calculate the form factors in the r.h.s. of the equations  (\ref{Gpi1}),
(\ref{Gpi6}), we use the modified impulse approximation (MIA) \cite{KrT21}
and, so, write them in terms of the integrals
\begin{equation}
	G^{(\pi)}_{1i0}(t) =\int d\sqrt{s} d\sqrt{s'}
	\varphi(s) G^{(0)}_{1i0}(s\,,t\,,s')\varphi(s')\;,
	\label{Gpi110}
\end{equation}
\begin{equation}
	G^{(\pi)}_{6k0}(t) =\int d\sqrt{s} d\sqrt{s'}
	\varphi(s) G^{(0)}_{6k0}(s\,,t\,,s')\varphi(s')\;,
	\label{Gpi610}
\end{equation}
Here
$i=1,4;k=1,4,6$;$\;G^{(0)}_{1i0}(s\,,t\,,s'),\,G^{(0)}_{6k0}(s\,,t\,,s')$
are components of the so-called free GFFs which desribe the system of
two free particles with total quantum numbers of pion
\cite{KrT21}, $\varphi(s)$ is the pion wave function in the sense of RQM
(\ref{phi}),
$s'\,,s$ are the invariant masses of the free two-particle system in the
initial and final states, respectively.

Now we have under integrals in
(\ref{Gpi110}) -- (\ref{Gpi610}) (compare \cite{KrT21}):
$$
G^{(0)}_{110}(s, t, s') = -\,\frac{R(s, t, s')\,(-t)}{\lambda(s,t,s')}
$$
$$
\times\left[(4M^2-t)\lambda(s,t,s') + \right.
$$
\begin{equation}
	+ \left.3\,t(s + s'- t)^2\right]\cos(\omega_1+\omega_2)
	\;,
	\label{G1100}
\end{equation}
$$
G^{(0)}_{140}(s, t, s') = -3\,M\,\frac{R(s, t, s')\,(-t)^2}{\lambda(s,t,s')}
$$
\begin{equation}
	\times\xi(s,t,s')(s+s'-t)\sin(\omega_1+\omega_2)\;,
	\label{G1400}
\end{equation}
$$
G^{(0)}_{610}(s, t, s') = \frac{1}{2}\,R(s, t, s')
$$
$$
\times\left[(s + s' - t)^2 - \right.
$$
\begin{equation}
	- \left.(4M^2 - t)\lambda(s,t,s')/(-t)\right]\cos(\omega_1+\omega_2)
	\;,
	\label{G6100}
\end{equation}
$$
G^{(0)}_{640}(s, t, s') = -\,\frac{M}{2}\,R(s, t, s')
$$
\begin{equation}
	\times\xi(s,t,s')(s+s'-t)\sin(\omega_1+\omega_2)\;,
	\label{G6400}
\end{equation}
$$
G^{(0)}_{660}(s, t, s') = R(s, t, s')
$$
\begin{equation}
	\times\lambda(s,t,s')\cos(\omega_1+\omega_2)\;,
	\label{G6600}
\end{equation}
where
$$
R(s, t, s') = \frac{(s + s'-t)}{2\sqrt{(s-4M^2) (s'-4M^2)}}\,
$$
$$
\times\frac{\vartheta(s,t,s')}{{[\lambda(s,t,s')]}^{3/2}}\;,
$$
$$
\xi(s,t,s')=\sqrt{-(M^2\lambda(s,t,s')+ss't)}\;,
$$
$\omega_1$ and $\omega_2$ are the Wigner spin-rotation parameters:
$$
\omega_1 =
\arctan\frac{\xi(s,t,s')}{M\left[(\sqrt{s}+\sqrt{s'})^2 - t\right] + \sqrt{ss'}(\sqrt{s} +\sqrt{s'})}\;,
$$
$$
\omega_2 = \arctan\frac{ \alpha (s,s') \xi(s,t,s')} {M(s + s' - t) \alpha (s,s') + \sqrt{ss'}(4M^2 - t)}\;,
$$
$\alpha (s,s') = 2M + \sqrt{s} + \sqrt{s'} $,
$\vartheta(s,t,s')= \theta(s'-s_1)-\theta(s'-s_2)$,
$\theta$ is the Heaviside function.
$$
s_{1,2}=2M^2+\frac{1}{2M^2} (2M^2-t)(s-2M^2)
$$
$$
\mp \frac{1}{2M^2} \sqrt{(-t)(4M^2-t)s(s-4M^2)}\;,
$$
$$
\lambda(a,b,c) = a^2 + b^2 +c^2 - 2(ab + ac + bc)\;.
$$

Recall that the form factors $G^{(0)}_{k0}(s,t,s'),\,k=1,6$ describe
gravitational features of a system of two particles without interaction.
Free two-particle form factors are regular generalized functions
(distributions)  given by the corresponding functionals, defined on the
space of test functions depending on the variables ($s,\,s'$). The
functionals, in turn, is a function of the variable $t$, a square of
momentum transfer. This variable is to be considered as a parameter.

In the frames of MIA the pion GFFs are functionals (\ref{Gpi1}) --
(\ref{Gpi610}), generated by the free two-particle GFFs of (\ref{T0}) on
test functions which are the products of the two-quark wave functions (see
((\ref{Gpi1}) -- (\ref{G6600})). There is a following difficulty in
calculating the form factor $G^{(\pi)}_{60}$ (an analogue of the form
factor $D$) in MIA.  Our expressions
(\ref{G6100}) -- (\ref{G6600}) show that $G^{(0)}_{60}\sim 1/t$ when $t\to
0$. A consequense of the mentioned singularity in $G^{(0)}_{60}$ is the
singularity in the pion form factor $G^{(\pi)}_{60}$ at $t\to 0$. So, for
correct description of the pion form factor $D$ we must overflow MIA and
add to the free form factor $G^{(0)}_{60}(s,t,s')$ an invariant term:
\begin{equation}
	G^{(0)}_{60}(s,t,s')\to G^{(0)}_{60}(s,t,s') + G^{(a)}_{60}(s,t,s')\;.
	\label{Ga}
\end{equation}

Strictly speakig, the calculation of this additional term should include
the construction of  a theory of interaction of graviton with two and more
quarks simultaneously. However, we avoid this difficult problem, choosing
a simple "minimal"$\,$ extension, based on the fact that the
non-relativistic limit of $G^{(0)}_{60}(s,t,s')$  is not singular. It is
important to note that the proximity of relativistic and non-relativistic
descriptions of GFFs at $t\to 0$ is largely adopted \cite{PoS18}. In fact,
in our "minimal"$\,$  variant the dynamical part of possible
additional contribution, the analog of meson exchange
currents in nuclear physics, remains out of scope.

The structure of the form factor $G^{(0)}_{60}(s,t,s')$, which is
the analog of form factor $D$ in the case of two free particles, is
defined by the following relation, obtained using (\ref{T0}), (\ref{Tq})
and (\ref{T0p1p2}):
$$
2\,t\,(A'_\mu A'^\mu)G^{(0)}_{60}(s,t,s') =
$$
$$ =
A\left\{\frac{1}{2}\left[g^{(u)}_{10}(t)+g^{(\bar d)}_{10}(t)\right]
\left[(\tilde K'_\mu \tilde K'^\mu)(A'_\mu A'^\mu) - \right.\right.
$$
$$
\left.- (\tilde K'_\mu A'^\mu)^2\right]\cos(\omega_1+\omega_2) +
$$
$$
+ \left[g^{(u)}_{40}(t)+g^{(\bar d)}_{40}(t)\right](\tilde K'_\mu A'^\mu)(\tilde R_\mu A'^\mu)\sin(\omega_1+\omega_2) +
$$
\begin{equation}
	+ \left. 2\left[g^{(u)}_{60}(t)+g^{(\bar d)}_{60}(t)\right]\,t\,(A'_\mu A'^\mu)\cos(\omega_1+\omega_2)\right\}\;,
	\label{GAA}
\end{equation}
where $\tilde K'_\mu,\,\tilde R_\mu$ -- 4-vectors  in (\ref{Tq}) appearing
after integration and summation in (\ref{T0p1p2}), $A=A(s,t,s')$ --
a multiplier defined by normalization of the Clebsh-Gordan coefficients
in (\ref{T0p1p2}).

The first two terms in r.h.s. have purely relativistic origin and are zero
in the non-relativistic limit, they do not depend on the quark form factors
$g^{(q)}_{60}$ from (\ref{Tq}) (an analogue of the quark form factor $D$)
and exactly these terms contain the
singularity. So, to advance in the simplest way, we require that the
additional term in  (\ref{Ga}) has zero for the non-relativistic limit
and contributes only to two first terms in (\ref{GAA}), compensating the
singularity. An important point, in our opinion, is the fact that the
mentioned dangerous terms depend on the vectors of one-particle
parametrization in r.h.s. of (\ref{T0p1p2}), (\ref{Tq}), while the
additional form factor in (\ref{Ga}), according to its meaning,
contributes to r.h.s. of (\ref{T0p1p2}), which contains the
one-particle currents. So, it seems natural that
$G^{(a)}_{60}(s,t,s')$ deforms the terms with the vectors of
parametrization of one-particle currents (\ref{kk'RG}), (\ref{GAA}).

We define the additional term in (\ref{Ga}) as to compensate the main
contribution of the relativistic series expansion of scalar products in
the first two terms in (\ref{GAA}). So, the divergent terms become closest
to their non-relativistic limit. The series expansion is carried out using
as parameters the quantities  $k/M,\; k'/M,\;\sqrt{(-t)}/M$ (the variable
$k$ is defined in (\ref{phi})). Thus we choose for  $G^{(a)}_{60}(s,t,s')$
the following relation:
$$
2\,(-t)\,(A'_\mu A'^\mu)G^{(a)}_{60}(s,t,s') =
$$
$$ =
A\left\{\frac{1}{2}\left[g^{(u)}_{10}(t)+g^{(\bar d)}_{10}(t)\right]
\left[\left[(\tilde K'_\mu \tilde K'^\mu)(A'_\mu A'^\mu) - \right.\right.\right.
$$
$$
\left.- (\tilde K'_\mu A'^\mu)^2\right] -
$$
$$
- \left.\left[(\tilde K'_\mu \tilde K'^\mu)(A'_\mu A'^\mu) - (\tilde K'_\mu A'^\mu)^2\right]_{lt}\right]\cos(\omega_1+\omega_2)+
$$
$$
+ \left[g^{(u)}_{40}(t)+g^{(\bar d)}_{40}(t)\right]\left[(\tilde K'_\mu A'^\mu)(\tilde R_\mu A'^\mu) - \right.
$$
\begin{equation}
	\left.- \left[(\tilde K'_\mu A'^\mu)(\tilde R_\mu A'^\mu)\right]_{lt}\right]\sin(\omega_1+\omega_2)\;,
	\label{Galt}
\end{equation}
where $[\ldots]_{lt}$ are the main terms of corresponding relativistic
series.

Because of the relations (\ref{Ga}) -- (\ref{Galt}),  only the main
terms of the relativistic expansion remain in diverging terms of
(\ref{GAA}) so making them finite. It is seen from the equations
(\ref{Ga}) -- (\ref{Galt}) that  the chosen scheme for going beyond MIA does
not contain arbitrariness. Taking into account the additional term
constracted with the use of (\ref{Ga}) -- (\ref{Galt}),  leads to the
following replacements in the equations (\ref{G6100}), (\ref{G6400}):
$$
(\tilde K'_\mu \tilde K'^\mu)(A'_\mu A'^\mu) - (\tilde K'_\mu A'^\mu)^2 =
$$
$$
= (4M^2-t)\lambda(s,t,s')/(-t) - (s + s'- t)^2\;\to\;
$$
\begin{equation}
	\;\to\; 4M^2\,(-t)\;,
	\label{sing1r}
\end{equation}
$$
(\tilde K'_\mu A'^\mu)(\tilde R_\mu A'^\mu) =
$$
\begin{equation}
	=	-\xi(s,t,s')(s + s'- t)\;\to\; -8M^2\,\tilde\xi(s,t,s')
	\label{big2r}
\end{equation}
$$
\tilde\xi(s,t,s')=
$$
\begin{equation}
	= M\sqrt{\left[(s'-4M^2)-\tilde s_1\right]\left[\tilde s_2 - (s'- 4M^2)\right]}\;,
	\label{txi}
\end{equation}
where
$$
\tilde s_1 = \left(\sqrt{s-4M^2} - \sqrt{(-t)}\right)^2\;,
$$
\begin{equation}
	\tilde s_2 = \left(\sqrt{s-4M^2} + \sqrt{(-t)}\right)^2\;.
	\label{ts12}
\end{equation}

After this procedure is carried out the purely relativistic terms of
(\ref{Gpi610}) take the form:
$$
G^{(\pi)}_{6k0}(t) =\int d\sqrt{s} d\sqrt{s'}
\varphi(s) G^{(R)}_{6k0}(s\,,t\,,s')\varphi(s')\;,
$$
\begin{equation}
	k=1,4\;.
	\label{Gpi610R}
\end{equation}
The new form factors $G^{(R)}_{6k0},\; k=1,4$ are obtained
using the equations (\ref{G6100}), (\ref{G6400}) and (\ref{sing1r}) --
(\ref{ts12}).
$$
G^{(R)}_{610}(s, t, s') = \frac{1}{2}\,\tilde R(s, t, s')
$$
\begin{equation}
	\times 4M^2\,t\cos(\omega_1+\omega_2)
	\;,
	\label{G6100R}
\end{equation}
$$
G^{(R)}_{640}(s, t, s') = -\,\frac{M}{2}\,\tilde R(s, t, s')
$$
\begin{equation}
	\times 8M^2\tilde\xi(s,t,s')\sin(\omega_1+\omega_2)\;,
	\label{G6400R}
\end{equation}
here $\tilde\xi(s,t,s')$ is defined by (\ref{txi}), and the functions
$\tilde R(s, t, s')$ and  $R(s, t, s')$ (in (\ref{G1100}) -- (\ref{G6600}))
differ in cutting function.
\begin{equation}
	\vartheta(s,t,s')\;\to\;\tilde\vartheta(s,Q^2,s')= \theta(s'-\tilde s_1)-\theta(s'-\tilde s_2)
	\label{ttheta}
\end{equation}

So,  to summarize, after the described minimal extension of MIA
is carried out, we calculate the pion
form factor $G^{(\pi)}_{60}$ using the equations (\ref{Gpi6}),
(\ref{Gpi610}), (\ref{G6600}), (\ref{Gpi610R}) -- (\ref{G6400R}).

\section{The model of quark gravitational structure and details of
calculation}
\label{sec: Sec 3}

In \cite{KrT21} we used the general method of the relativistic
invariant parametrization of the matrix elements of the local operators
established in Ref. \cite{ChS63} for
systems with arbitrary spin. Although now we consider actually the pion, we
preserve for reasons of convenience the notations of the preceding
paper. The pion GFFs in this parametrization are connected
with commonly used (see, e.g., \cite{PoS18}) by the following relations:
\begin{equation}
        A^{(\pi)}(t) = G^{(\pi)}_{10}(t)\;,\quad
        D^{(\pi)}(t) = -2\,G^{(\pi)}_{60}(t)\;,
        \label{ADpiG16}
\end{equation}
where $t = (p_\pi - p_\pi ')^2$, and $\;p_\pi ', p_\pi$
are the pion 4-momenta in the initial and the final states, respectively,
and $G^{(\pi)}_{10},\;G^{(\pi)}_{60}$ are given by the equations
(\ref{Gpi1}) -- (\ref{G1400}), (\ref{G6600}), (\ref{Gpi610R}) --
(\ref{G6400R}).


The quark GFFs in our approach are connected with GFFs that are commonly
used for particles with spin $1/2$ (see, e.g., \cite{PoS18}) by the
following relations \cite{KrT21}:
$$
g^{(q)}_{10}(t) = \frac{1}{\sqrt{1-t/4M^2}}\left[\left(1 -\frac{t}{4M^2}\right)\right.
A^{(q)}(t) +
$$
\begin{equation}
        + \left. 2\frac{t}{4M^2}J^{(q)}(t)\right]\;,
        \label{g10}
\end{equation}
\begin{equation}
        g^{(q)}_{40}(t) = -\,\frac{1}{M^2}\frac{J^{(q)}(t)}{\sqrt{1 - t/4M^2}}\;,
        \label{g40}
\end{equation}
\begin{equation}
        g^{(q)}_{60}(t) = -\,\frac{1}{2}\sqrt{1 - \frac{t}{4M^2}}D^{(q)}(t)\;.
        \label{g60}
\end{equation}

We assume that the GFFs of $u$- and $\bar d$-quarks
are equal: $g^{(u)}_{i0}(t) = g^{(\bar d)}_{i0}(t)\;, i=1,4,6$.

To define the explicit  form of the quark GFFs we recall our
calculation of the electroweak structure of the pion. We derived
the functional form
of the electromagnetic form factor of the quark from the behavior of our
charge form factor of pion at $(-t)\to\infty$ \cite{KrT01},
which turned out to coincide with that of QCD, as mentioned above:
\begin{equation}
        f_q(t) = \frac{1}{1 + \ln\left(1 - \langle r^2_q\rangle t/6\right)}\;,
        \label{fq}
\end{equation}
where $\langle r^2_q\rangle$ is a mean square charge radius of the
constituent quark.
The charge form factor of the constituent coincides with
the function (\ref{fq}) and the magnetic form factor is equal
to this function multiplied by the corresponding magnetic moment
\cite{KrT01}. Now we admit a similar definition for the quark GFFs
(\ref{g10}) -- (\ref{g60}) in term of (\ref{fq}):
$$
A^{(q)}(t)=f_q(t)\;,\quad J^{(q)}(t) = \frac{1}{2}f_q(t)\;,
$$
\begin{equation}
        D^{(q)}(t) = D_q\,f_q(t)\;,
        \label{AJDfq}
\end{equation}
where $D_q$ is the $D$-term of the constituent quark. These equations give
standard static limits
(see, e.g., \cite{PoS18}):
\begin{equation}
        A^{(q)}(0)=1\;,\quad J^{(q)}(0) = \frac{1}{2}\;,\quad D^{(q)}(0) = D_q\;,
        \label{Ag0}
\end{equation}
We set the parameter $\langle r^2_q\rangle$ in (\ref{fq}),
to be equal to the mass  MSR of the quark and to define the slope of the
quark form factor $D$ at zero. The corresponding
actual value is given by the following form \cite{VoL90, PoH90, TrT94,
CaG96}:
\begin{equation}
        \langle r^2_q\rangle  \simeq \frac{0.3}{M^2}\;.
        \label{rq}
\end{equation}

For the calculation of the pion GFFs according to (\ref{Gpi1}) -- (\ref{Gpi610}),
(\ref{Gpi610R}), (\ref{ADpiG16}) we use in (\ref{phi}) consequently one of the
following model wave functions:
\begin{equation}
        u(k) = 2\left(1/(\sqrt{\pi}\,b^3)\right)^{1/2}\exp\left(-\,k^2/(2\,b^2)\right)\;,
        \label{wfHO}
\end{equation}
\begin{equation}
        u(k) = 16\left(2/(7\pi\,b^3)\right)^{1/2}\left(1 + k^2/b^2\right)^{-3}\;,
        \label{wfPL3}
\end{equation}
\begin{equation}
        u(k) = 4\left(2/(\pi\,b^3)\right)^{1/2}\left(1 + k^2/b^2\right)^{-2}\;.
        \label{wfPL2}
\end{equation}
We use the parameter $b$ in (\ref{wfHO}) -- (\ref{wfPL2}) fixed previously
in the works \cite{KrT21, KrT01, Kru97} on the electroweak
properties of the pion. The different type of functions correspond to the
different type of confinement. So, in the model (\ref{wfHO}) quadratic
confinement is carried out, the model (\ref{wfPL3}), as our calculations
of the electroweak properties of a pion in the work \cite{KrT01} show, gives
predictions very close to the model with linear confinement \cite{Tez91},
and finally, the wave function (\ref{wfPL2}) corresponds to a confinement
weaker than a linear one. We calculate the pion GFFs using the same
value of the constituent-quark mass, $M$= 0.22 GeV, as in our previous
works (see, e.g., \cite{KrT01}).

So, now only one free parameter remains -- the $D$-term  of the constituent
quark $D_q$ in the form factor (\ref{AJDfq}), (\ref{Ag0}). In Sect.
\ref{sec: Sec 4} we fix this free parameter and present the gravitational
properties of pion calculated taking into account the constituent-quark
gravitational structure.

\section{The calculation of the pion GFFs and
their static moments}
\label{sec: Sec 4}

\begin{figure}[h!]
        \epsfxsize=0.9\textwidth
        \centerline{\psfig{figure=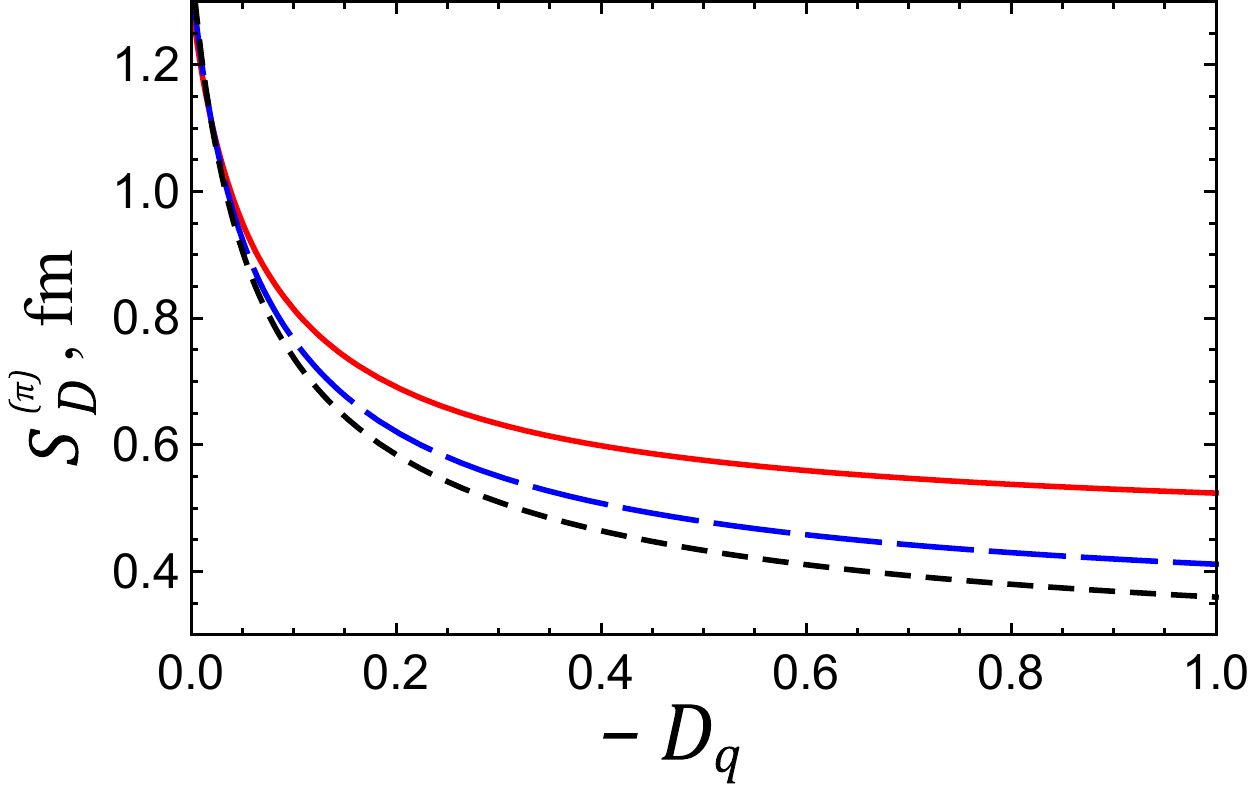,width=9cm}}
        \vspace{0.3cm}
        \caption{The  slope at zero of the normalized to pion $D$-term
                 form factor $D$ of pion as a function of the $D$-term of
                 the constituent quark $D_q$ calculated with various model
                 wave functions (\ref{wfHO})  -- (\ref{wfPL2}). Full line
                 (red) -- with wave function (\ref{wfHO}); dashed  line
                 (blue) -- with (\ref{wfPL3}), short-dashed line (black) --
                 with (\ref{wfPL2}).}
\label{fig:1}
\end{figure}

To calculate numerically the pion GFFs we need, first of all, to
fix the remaining free parameter $D_q$ which enters the pion
form factor $D$. Namely, we use the static moments of $D^{(\pi)}(t)$. For
the time being there is no strict experimental estimation of the value
$D^{(\pi)}(0)$ \cite{PoS18}. However, the first data for the
slope at zero of the normalized to pion $D$-term form factor $D$ of
pion extracted from the process
$\gamma^*\gamma\to \pi^0\pi^0$ were presented in Ref. \cite{KuS18}:
\begin{equation}
        S^{(\pi)}_D = \left(0.82 - 0.88\right)\,\hbox{fm}\;.
        \label{r2mexp}
\end{equation}
We use this estimation to fix $D_q$ taking into account the following
definition \cite{PoS18}:
\begin{equation}
        \left(S^{(\pi)}_D\right)^2 =
        \frac{6}{D^{(\pi)}(t)}\,\left.\frac{dD^{(\pi)}}{dt}\right|_{t=0}\;.
        \label{r2mDpi}
\end{equation}
As all but one parameters of the model were fixed previously in the works
on electroweak structure of the pion, then
$S^{(\pi)}_D$ (\ref{r2mDpi}) is a function of the
$D$-term of the constituent quark, $D_q$, only. This function
is given in Fig. \ref{fig:1} for three types of model quark interaction in
pion (\ref{wfHO}) -- (\ref{wfPL2}).

The results presented in the figure show that for each of the model wave
functions there is an interval of the values of  $D_q$ which gives the
interval  (\ref{r2mexp}) of the values of
$S^{(\pi)}_D$. Moreover, there exists the interval
of the variable $D_q$, for which $S^{(\pi)}_D\,$
(\ref{r2mDpi}) falls in the interval (\ref{r2mexp}) for all the wave
functions (\ref{wfHO})--(\ref{wfPL2}), without any additional variation of
the parameters. This interval is
\begin{equation}
        D_q = -\left(0.0715 - 0.0709\right)\;.
        \label{Dqmech}
\end{equation}
Note that this interval is rather narrow ($<0.5$\%). The existence of this
interval recalls the fact that when describing the electroweak properties
 of pion in RQM we obtained the same values of the physical
characteristics of the constituent quark for all model wave functions
\cite{KrT01}. Now we add the quark $D$-term to the previous set (the mass,
the charge MSR, the anomalous magnetic moment).

Now, with (\ref{Dqmech}), all the parameters are fixed and we do
calculate the pion GFFs. Note that the pion form factor $A$,
(\ref{ADpiG16}), (\ref{Gpi1}), does not depend on $D_q$ and is determined
by the parameters obtained previously, so, it is predicted. The standard
condition $A^{(\pi)}(0) = 1$ is fulfilled automatically for arbitrary
values of parameters, if the quark form factors satisfy Eq. (\ref{Ag0}).

The results of calculation of the static gravitational moments of
the pion are presented in the Table \ref{tab:table1}. The mass
MSR is defined following the relation \cite{PoS18}:
\begin{equation}
        \langle r^2\rangle_{mass} =
 6\,\frac{1}{A^{(\pi)}(t)}\left.\frac{dA^{(\pi)}}{dt}\right|_{t=0}\;.
\label{r2mass}
\end{equation}

In the last line of the table we present the results for the pion
mechanical radius using the standard expression
\begin{equation}
	\langle r^2\rangle_{mech} = 6\frac{D^{(\pi)}(0)}{\int_{-\infty}^0\,D^{(\pi)}(t')dt'}\;,
	\label{rmech}
\end{equation}
For the calculation it is important to have the form factor $D$ in the
large range of momentum transfer including its asymptotics. This is a
complicated problem, which is out of scope of this paper. It
will be considered elsewhere. A thorough correct full-fledged approach can
be realized in the spirit of papers \cite{KrT98, TrT13, KrT17, TrT21}.
The GFF obtained in the present paper  predictably gives
underestimated values of the mechanical radius as compared with mass and
electromagnetic radii. We believe that the detailed calculation which is in
progress will give the ultimate truth.

Note that the deviations of the values $S^{(\pi)}_D$
from that given in the Table \ref{tab:table1} when the parameter
$D_q$ is changed in the interval (\ref{Dqmech}) is approximately  0.1\%.
\begin{table}
        \caption{\label{tab:table1}. The results of calculating the
        static gravitational moments of the pion with different model
         wave functions (\ref{wfHO}) - (\ref{wfPL2}). The mass of
         constituent quarks is $M = 0.22$ GeV, their MSR is given in
         (\ref{rq}) and the $D$-term $D_q$ is taken from (\ref{Dqmech}).
         The deviation from the given values of
         $S^{(\pi)}_D$ when the parameter $D_q$ is
         varied in the interval (\ref{Dqmech}) is about 0.1\%.}
\begin{tabular}{lccc}
	\hline
	\hline
	Model                                 &(\ref{wfHO})     &(\ref{wfPL3})    &(\ref{wfPL2})  \\
	\hline
	$b$, GeV                              &0.3500	        &0.6131           &0.4060          \\
	$A^{(\pi)}\,'(0)$, GeV$^{-2}$         &1.074            &1.134            &1.217           \\
	$\langle r^2\rangle^{1/2}_{mass}$, fm &0.50             &0.52             &0.53            \\
	$-D^{(\pi)}(0)$                       &0.704--0.700     &0.653--0.649     &0.620--0.617    \\
	$-D^{(\pi)}\,'(0)$,GeV$^{-2}$         &2.331--2.327     &1.978--1.976     &1.773--1.771    \\
	$S^{(\pi)}_D$, fm                     &0.88             & 0.84            &0.82            \\
	$\langle r^2\rangle^{1/2}_{mech}$, fm &0.162--0.163     &0.153--0.154     &0.133--0.134    \\
	\hline
	\hline
\end{tabular}
\end{table}

Let us do some remarks about our values of the pion $D$-term. As is well
known, in theories with broken chiral symmetry (see, e.g.,
\cite{VoZ80, NoS81}) the pion $D$-term is equal to $-1$. However, if
EMT contains only the contribution of quarks, without taking gluons
into account, then its value is approximately $\simeq-0.75$
obtained
in Ref. \cite{KuS18} in agreement
with the soft pion theorem $D = -1$, given that the extracted value does
not include the gluon contribution \cite{PoS18,Pas18}

Our values given in the Table \ref{tab:table1} are close to this value.
The same is true about the comparison of our slope at zero of the pion
$D$-form factor with that of chiral theory, where
$-D^{(\pi)}\,'(0) = 2.40\;\hbox{GeV}^{-2}$. The results for pion GFFs
satisfy the well-known relation (see, e.g., \cite{PoS18})
$-D^{(\pi)}\,'(0) > A^{(\pi)}\,'(0)$. In addition, our values of
$A^{(\pi)}\,'(0)$ are close to the estimations given, for example, in the
same review:
\begin{equation}
        A^{(\pi)}\,'(0) = \left(1.33-2.02\right)\,\hbox{GeV}^{-2}\;.
        \label{Api0c}
\end{equation}

Let us compare the mass and charge MSRs of the pion.
The experimental pion charge MSR is (see, e.g., \cite{KuS18}):
\begin{equation}
        \sqrt{\langle r^2\rangle_{ch}} = 0.672\pm 0.008\,\hbox{fm}\;.
         \label{r2chd}
\end{equation}
The mass MSR was estimated in  \cite{KuS18} and the following interval of
the values was obtained:
\begin{equation}
        \sqrt{\langle r^2\rangle_{mass}} = \left(0.32-0.39\right)\,\hbox{fm}\;.
        \label{r2massd}
\end{equation}

Our result for the mass radius (see Table \ref{tab:table1}) lays
much closer to the charge radius (\ref{r2chd}) than
(\ref{r2massd}). Curiously enough, our difference between the
mass and the charge radii is the same as obtained  in
the work  \cite{WaK21PRD} for the proton
($\Delta R_{cm} = 0.1709\pm 0.0304$ fm).
We recall that our result for $\langle r^2\rangle^{1/2}_{mass}\,$
(\ref{r2mass}) is a direct consequence of our model of electroweak
properties of the pion without new fittings.

\begin{figure}[h!]
        \epsfxsize=0.9\textwidth
        \centerline{\psfig{figure=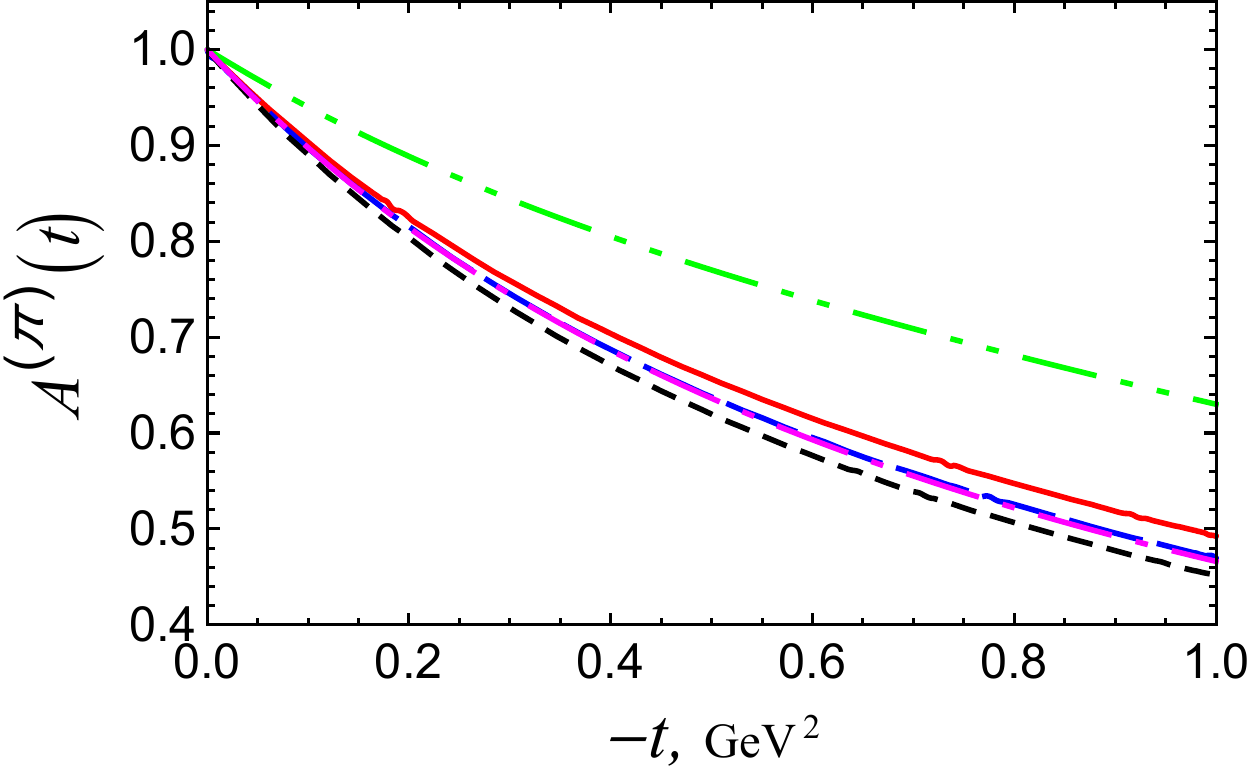,width=9cm}}
        \centerline{\psfig{figure=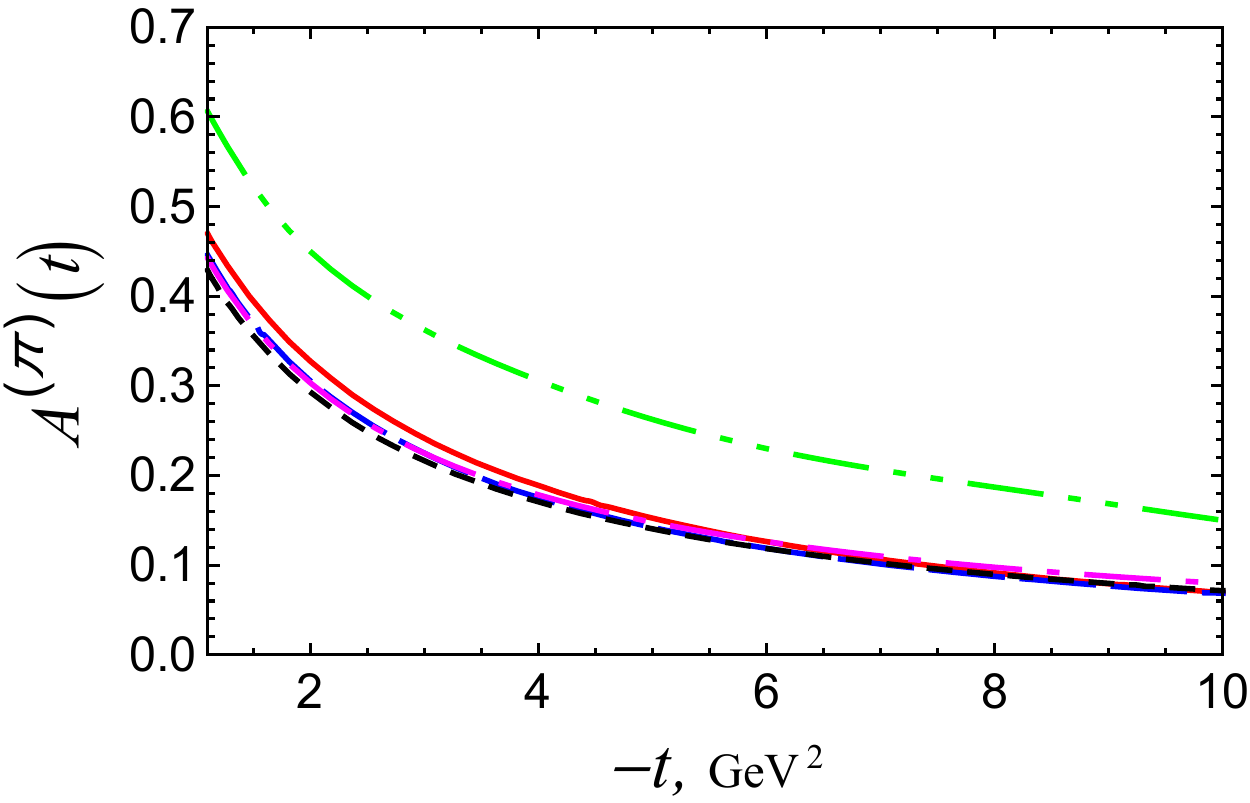,width=9cm}}
        \vspace{0.3cm}
        \caption{The pion form factor $A$ calculated for the  quark
structure defined by (\ref{g10}) -- (\ref{rq}) and different model wave
                functions (\ref{wfHO}) -- (\ref{wfPL2}) with
                $b$ given in Table \ref{tab:table1}, $M=$0.22 GeV.
                Full line (red) -- with wave function (\ref{wfHO}); dashed
                line (blue) -- with (\ref{wfPL3}), short-dashed line
                (black) -- with (\ref{wfPL2}), dotted dashed line
                (magenta) -- the results for the fitting (\ref{Fit});
                double-dotted dashed line (green) presents the result of
                the extraction of the form factor $A$ given in the paper
                \cite{KuS18}.}
\label{fig:2}
\end{figure}

\begin{figure}[h!]
        \epsfxsize=0.9\textwidth
        \centerline{\psfig{figure=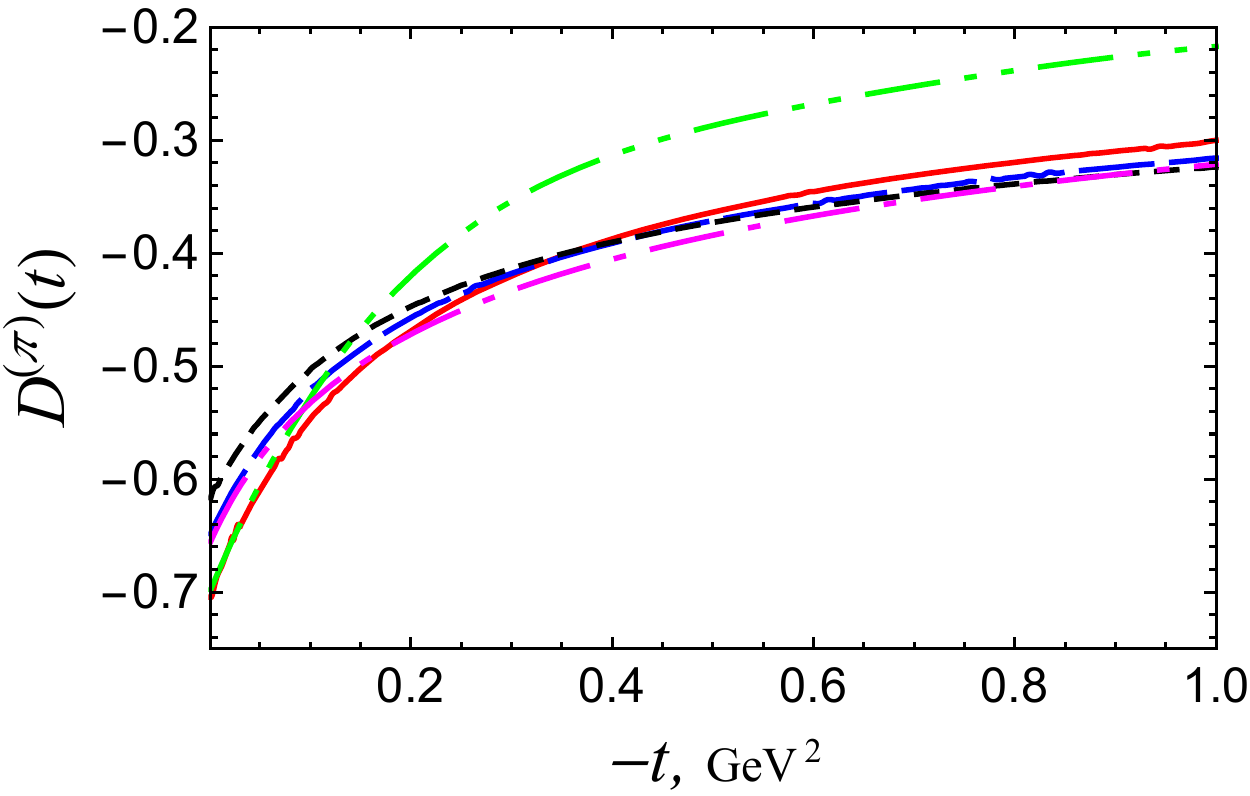,width=9cm}}
        \centerline{\psfig{figure=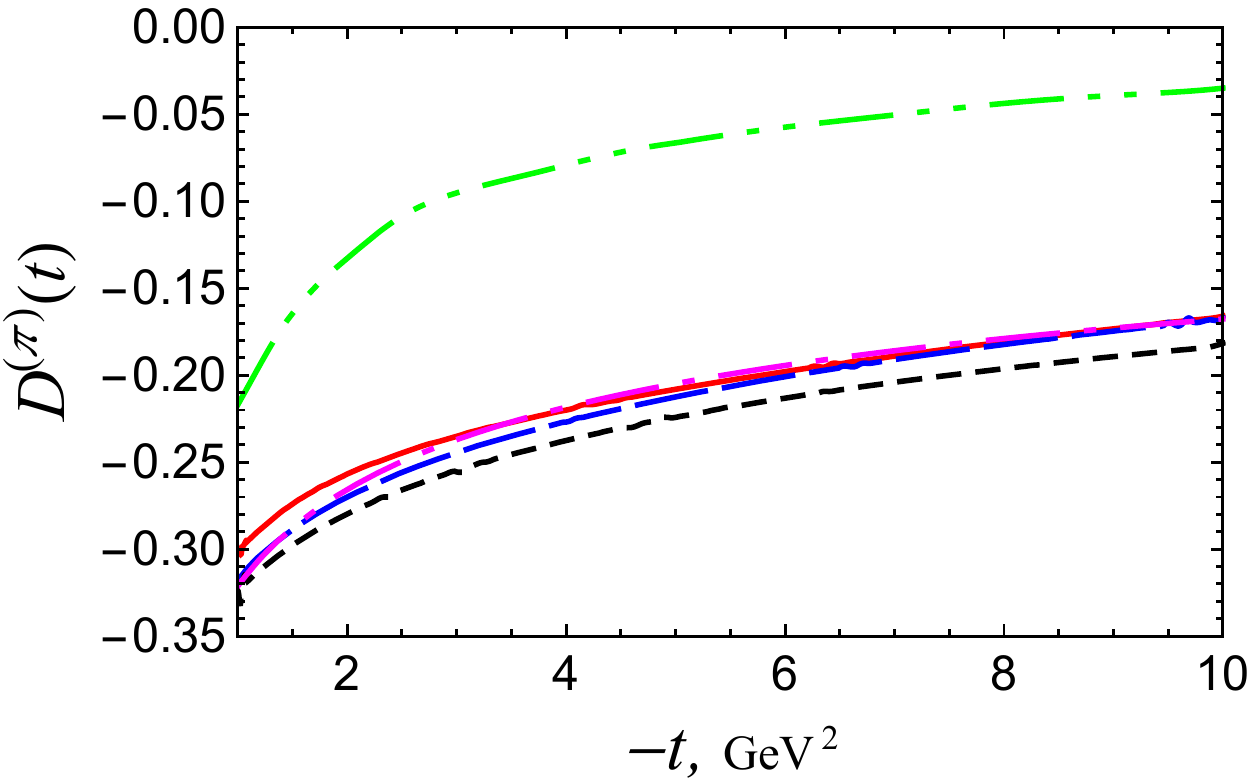,width=9cm}}
        \vspace{0.3cm}
        \caption{The pion form factor $D$
(\ref{ADpiG16}), (\ref{Gpi6}), (\ref{Gpi610}), (\ref{Gpi610R})
calculated for the  quark
structure defined by (\ref{g10}) -- (\ref{rq}) and different model wave
                functions (\ref{wfHO}) -- (\ref{wfPL2}) with
                $b$ given in Table \ref{tab:table1}, $M=$0.22 GeV.
                 Two bound values of $D_q$ from the interval
                 (\ref{Dqmech}) were used for the calculation, but the
                obtained curves are unseparable.
                Full line (red) -- with wave function (\ref{wfHO}); dashed
                line (blue) -- with (\ref{wfPL3}), short-dashed line
                (black) -- with (\ref{wfPL2}), dotted dashed line
                (magenta) -- the results for the fitting (\ref{Fit});
                double-dotted dashed line (green) presents the result of
                the extraction of the form factor $D$ given in the paper
                \cite{KuS18}.}

\label{fig:3}
\end{figure}

The results of calculation of the pion form factors $A$ and $D$ at low and
intermediate momentum transfers for the wave functions
(\ref{wfHO}) -- (\ref{wfPL2}) and the quark structure (\ref{g10}) -- (\ref{rq})
are presented in  Figs. \ref{fig:2} and \ref{fig:3}.
For comparison, the results for GFFs as extracted \cite{KuS18} from the
experimental data are given, too. Note that the slope of our curves is
close to that of the experimental one at large $Q^2$.
The results
for the lower and higher bounds of $D_q$ from (\ref{Dqmech}) are
presented, but in fact can not be separated from one another in the
figures.

The pion GFFs, shown in figures \ref{fig:2} and \ref{fig:3}, calculated
for the values of the quark $D$-term, $D_q$, in the interval (\ref{Dqmech}),
form a rather narrow bunch corresponding to the quark mass $M=0.22$ GeV.
The curves remain agglomerated in the bunch up to squares of momentum
transfer $\sim 10\;\hbox{GeV}^2$. The schema of bunches works anew.

As the dependence of the pion GFFs on the choice of model wave
functions  (\ref{wfHO}) -- (\ref{wfPL2})
is weak, it seems us reasonable to construct an averaged fit for these
form factors. We take  the fitting function in the following form, which
is used, for example in \cite{PaP14}:
\begin{equation}
	F(t) = \frac{F(0)}{(1- t/a)^n}\;.
	\label{Fit}
\end{equation}
Least square fitting gives the following values of parameters. For the
form factor $A^{(\pi)}$ we obtain: $F(0)=1$, $\;a =
0.882\;\hbox{GeV}^{2}$,$\;n = 1.007$ and for the form factor $D^{(\pi)}$:
$F(0) = -0.657$, $\;a = 0.0949\;\hbox{GeV}^{2}$, $\;n = 0.293$. The value
of the parameter $F(0)$ for $D^{(\pi)}$ is equal to average over models
value of the pion $D$-term (see Table \ref{tab:table1}) and the slopes of
the formfactors $A$ and $D$ at zero -- to the corresponding averaged
slopes. Obviously, there is no such correspondence for the pion
mechanical radius (see the remarks above, after (\ref{rmech})). The
results of the fitting are shown in figures \ref{fig:2} and \ref{fig:3}.

\section{Conclusions}
\label{sec: Sec 5}

In this work we extend our relativistic theory of
composite-particle systems which describe simultaneously their
electroweak and gravitational properties to calculate the gravitational
characteristics of the pion. The approach is based on a version of the
instant-form relativistic quantum mechanics. In the preceding work
\cite{KrT21} we formulated the mathematical base and described the
principal schema of deriving the gravitational characteristics in the
approach. In the present work we give a detailed technique of calculation,
thus transforming the approach into a quantitative  method. The extension
includes three main points: the constituent quarks are no longer
point-like, we analyze different types of the quark interaction in pion,
and we formulate a minimal way to overstep the frame of our
modified impulse approximation (MIA) for correct description of the pion
form factor $D$.

We derive the equations for the pion GFFs, the mass radius and the slope
at zero of the normalized to pion $D$-term  form factor $D$ of pion
in term of parameters describing the quark structure.
 Most of the parameters have been fixed even earlier
in our works on the pion electromagnetic form factors. The only free
parameter is the $D$-term of the constituent quark, which we fix by
fitting the result for the slope
at zero of the normalized to pion $D$-term  form factor $D$ of pion, to a
choosen experimental value  \cite{KuS18}. The results for the  form factor
$A$ and for the mass MSR of pion do not depend on this new parameter and,
so, are direct predictions of our approach. As a whole, the obtained now
results for the gravitational characteristics of pion can be considered as
firm predictions if calculated with a new future value of the
mentioned slope.

Using these parameters we calculate the pion GFFs,  their derivatives and
static moments. We present the pion GFFs
in the range of  low and intermediated momentum transfers up to $\sim
10\;\hbox{GeV}^2$. At large $Q^2$ the slope of our curves is  close to that
of \cite{KuS18}. The curves for different model quark-antiquark wave
functions are agglomerated to form bunches corresponding to the quark
mass $M=0.22$ GeV, as it is in the electroweak case. The fits of the
calculated pion GFFs, averaged over obtained with three different model
wave functions, are constructed. The values of the static moments, mass MSR
and the $D$-term of pion are consistent with the values given in the
literature.

It seems obvious that  because of extreme weakness of the
gravitational interaction at hadron scale, the information on  GFFs will
be extracted from electroweak processes as yet. So, we believe that our
theory which describes electroweak and gravitational properties
simultaneously, is based on the unique foundations, and exploits common
model parameters, will be useful.
Moreover, it is possible that our unified approach  can construct another
bridge between electroweak and gravitational properties of composite
systems, complementary to GPD.

The work was supported in part by the Ministry of Science and Higher
Education of the Russian Federation (grant N 0778-2020-0005)(A.K.).

\end{document}